\documentclass[twocolumn, trackchanges]{aastex701} %, linenumbers
\definecolor{trackchange}{cmyk}{0,1.0,0.6,0.2}

\usepackage{multirow}

\usepackage{graphics,epsf}
\usepackage[utf8]{inputenc}
\usepackage{amsmath}                % American Mathematical Society package
\usepackage{amsfonts}               % American Mathematical Society fonts
\usepackage{amssymb}                % American Mathematical Society symbol
\usepackage{epsfig}                 % EPS figures
\usepackage{graphicx}               % Required for inserting images
\usepackage{float}
\usepackage{color}
\usepackage{multirow}               % double row table entries

\hypersetup{
    colorlinks=true,
    linkcolor=red,   
    urlcolor=cyan}

\usepackage[colorinlistoftodos]{todonotes}

\newcommand{\km}{{~\rm km}}
\newcommand{\s}{{~\rm s}}

\newcommand{\g}{{~\rm g}}
\newcommand{\G}{{~\rm G}}
\newcommand{\K}{{~\rm K}}
\newcommand{\erg}{{~\rm erg}}

\begin{document}

\title{Identifying a jet-shaped point-symmetric morphology in the supernova remnant HB9}

\author[0000-0003-0375-8987]{Noam Soker}
\affiliation{Department of Physics, Technion - Israel Institute of Technology, Haifa, 3200003, Israel; 
soker@technion.ac.il}
\email{soker@technion.ac.il}

\begin{abstract}
I identify a point-symmetric morphology in the supernova remnant HB9 (G160.9+2.6), supposedly a core-collapse supernova (CCSN) remnant (CCSNR), and attribute this structure to two or more pairs of jets that exploded the star within the jittering-jets explosion mechanism (JJEM). The pairs of jets that shape the point-symmetric morphology of HB9 are part of a larger number of pairs that exploded the star. The point-symmetric structure is composed of a pair of ears, a pair of ears/nozzles, a pair of dents, and a pair of bright bars. Although these pairs of morphological features are small and the point-symmetric structure is subtle, I consider the identification robust. I use visual inspection to identify the point-symmetric morphology of HB9, a method commonly used to identify jet-shaped, point-symmetric morphologies in other astrophysical objects, such as cooling-flow clusters of galaxies and planetary nebulae. I present arguments for jet-driven shaping rather than CSM shaping. These findings support earlier claims that the JJEM is the primary explosion mechanism of CCSNe. 
\end{abstract}


% ==================================
\section{Introduction} 
\label{sec:intro}
% ==================================

There are two heavily studied theoretical core-collapse supernova (CCSN) explosion mechanisms that compete in recent years for being the primary explosion mechanism of CCSNe. Both mechanisms utilize the gravitational energy released by the collapsing core to explode the star.  
 
In the delayed neutrino-driven (neutrino-heating) explosion mechanism (\citealt{BetheWilson1985_OGdelayedneutrino};
and some papers since 2026: \citealt{Akaho2026, Andresenetal2026, Burrowsetal2026, Calvertetal2026, ChenCHetal2026, EggenbergerAndersenetal2026,  LuoZhaKajino2026, Mezzacappa2026, Neopaneetal2026}), neutrinos emitted by the cooling proto-neutron star (NS) heat the material behind the stalled shock, revive the stalled shock, and explode the star. 
In the Jittering jets explosion mechanism (JJEM; \citealt{Soker2010, PapishSoker2011}), and some papers since 2026 \citealt{Soker2026DustJets, Soker2026Failed, WangShishkinSoker2026, KlimovSoker2026, ShishkinSoker2026SNR0540, AkashiSoker2026BG11, BraudoSoker2026Pipe, ShishkinAkashiSoker2026}), pairs of jets with varying axes explode the star. 
The magnetorotational mechanism, where the NS launches one pair of jets with a fixed axis, requires rapidly rotating pre-collapse core, and, therefore, can at best explain a small fraction of all CCSNe  (e.g., \citealt{Shibataetal2025, Mannoetal2026, PanLi2026, Griffithsetal2026} for some recent papers); if the NS collapses to a black hole that launches the pair of jets, this is a collapsar (e.g., \citealt{BoppGottlieb2025, Gottliebetal2025}). On the large differences between the magnetorotational mechanism and the JJEM, and on the relation of the neutrino-driven mechanism and the JJEM to other energy sources, such as magnetars, see \cite{Soker2025G11}.  

At present, observations of CCSNe at and shortly after the explosion cannot robustly distinguish between the two alternative explosion mechanisms. For example, studies (e.g., \citealt{Kangasetal2026}) attribute CCSNe with explosion energies of $>2 \times 10^{51} \erg$ that the neutrino-driven cannot explain to rare cases of a pairs of energetic jets (like the magnetorotational mechanism); but these are rare cases that do not testify on the majority of CCSNe. 
This leaves the morphologies of CCSN remnants (CCSNRs) as the main observable that can robustly decide between them (e.g., \citealt{Soker2024UnivReview, Soker2025Learning}). 
According to the JJEM, the pairs of jittering jets shape a large fraction of CCSNe to possess point-symmetric morphologies. These morphologies are systematically produced in recent three-dimensional (3D) simulations of the JJEM (e.g., \citealt{Braudoetal2025, Braudoetal2026, AkashiSoker2026a, AkashiSoker2026BG11, BraudoSoker2026Pipe, BraudoSoker2026Precess, ShishkinAkashiSoker2026}). 
In contrast, neutrino-driven and magnetorotational mechanisms do not explain point-symmetric structures in CCSNRs; they can explain non-spherical geometries, such as dipole and even axial symmetry (e.g., \citealt{Wessonetal2026, Giudicietal2026}), but not point-symmetric morphologies.

For these, any new CCSNR identified with a point-symmetric structure is a significant contribution to our quest to reveal and understand the explosion mechanism of CCSNe. 

In the last five years, starting with SNR 0540-69.3 (\citealt{Soker2022SNR0540, ShishkinSoker2026SNR0540}), studies have identified point-symmetric morphologies in about 20 CCSNRs and attributed these morphologies to the JJEM. A very recent addition is SNR G7.7-3.7, where \cite{Luoetal2026G77} identified two pairs of opposite ears (which they termed blowouts), which \cite{Soker2026Long} attributed to the JJEM. In between, studies identified point-symmetric morphologies and attributed them to the JJEM in many other CCSNRs, including the three iconic and most studied CCSNRs: SN 1987A (\citealt{Soker2024NA1987A, Soker2024Keyhole, Soker2024PNSN, Soker2026SN1987Amulecular, Soker2026SN1987AFailed}), Cassiopeia A (\citealt{BearSoker2025, ShishkinAkashiSoker2026}), and the Crab Nebula (SN 1054; \citealt{ShishkinSoker2025Crab, Soker2026DustJets, Soker2026RNAASCrab}).

In this study, I identify a subtle point-symmetric structure in the SNR HB9 (SNR G160.9+2.6; HBH 9). SNR HB9 has been observed in several wavebands (e.g., \citealt{AbraninBazelian1982HB9, LeahyAschenbach1995HB9, LeahyTian2007HB9, Araya2014HB9, Sezeretal2019HB9, Saitoetal2020HB9, Erginetal2021HB9, Okaetal2022HB9, LiJetal2024HB9, Xiao2025HB9}). (For some other studies of HB9, see, e.g., \citealt{Leahyetal1998HB9, Leahyetal1999HB9, Gosachinskii2013HB9, XiaoLetal2026HB9}). In this study, I deal only with the morphology of HB9. In Section \ref{sec:HB9}, I identify a point-symmetric structure in HB9. 
 In Section \ref{sec:Extension}, I discuss the emission south of the main CCSNR and suggest it might be related to the jets that powered and shaped CCSNR HB9. 
 In Section \ref{sec:CSM}, I compare shaping by a CSM and by jets in the explosion process: I discuss HB9 and the recent dispute over the shaping process of the ears of SNR S147. I summarize this study, further supporting the JJEM, in Section \ref{sec:Summary}.

% =========================
\section{Identifying a point-symmetric morphology in HB9}
\label{sec:HB9}
% =========================

\begin{figure*}
\centering
\includegraphics[trim=0.0cm 5.3cm 0.0cm 0.0cm ,clip, scale=0.75]{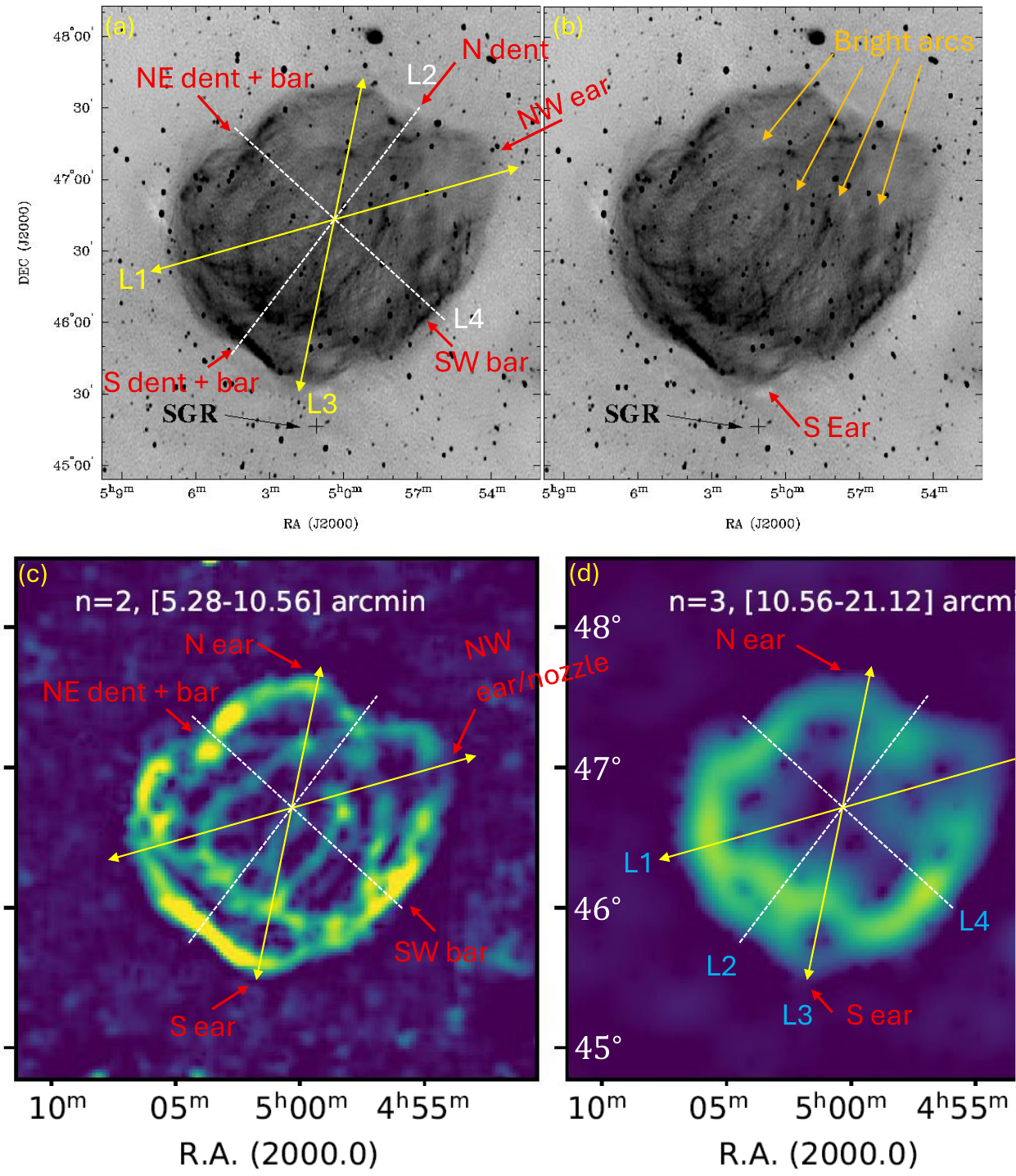} 
\caption{The two upper panels are identical: an 1.4 GHz radio image of HB9 adapted from \cite{GaenslerChatterjee2008}; see also \cite{LeahyTian2007HB9}. SGR is the magnetar SGR 0501+4516, which is unrelated to HB9 (e.g., \citealt{Chrimesetal2025}). In the left one, I added a wind rose with two dashed lines and two double-sided arrows. These 4 lines connect opposite structure features and define the point-symmetric morphology of HB9. 
Lower panels: radio images from \cite{Xiao2025HB9} of the FAST subband at 1085 MHz; see details of the images there. I added the wind rose from panel (a). 
}
\label{fig:HB9radioFigure}
\end{figure*}

\begin{figure}
\centering
\includegraphics[trim=0.0cm 16.8cm 0.0cm 2.1cm ,clip, scale=0.75]{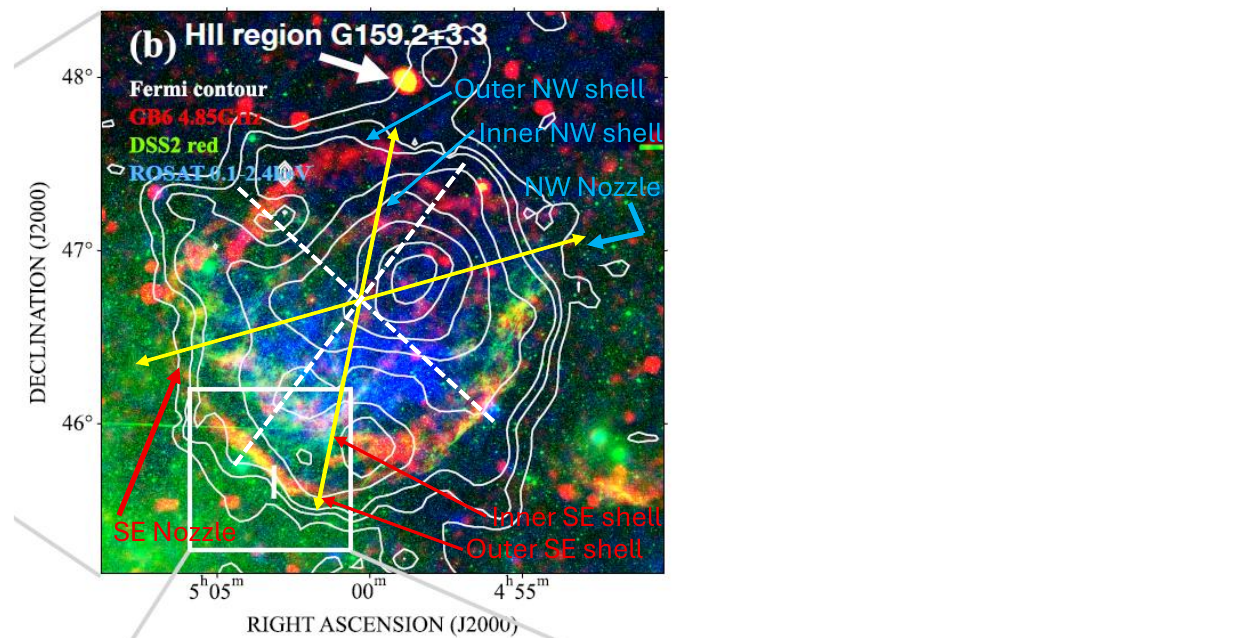} 
\caption{  Multi-wavelength images of the SNR HB9, the H II region G159.2+3.3, and the surrounding area. An image adopted from \cite{LiJetal2024HB9}: Red: 4.85 GHz radio
continuum image from the Green Bank 6-cm (GB6) survey (\citealt{Gregoryetal1996}); Green: DSS2 red-band image that covers the H$\alpha$ line; Blue: ROSAT/PSPC broad-band (0.1–2.4 keV) image from the ROSAT All Sky Survey  (\citealt{Vogesetal1999}).
The contours outline the Fermi-LAT image. I added the wind rose from Figure 1 and marked the NW and SE nozzles. I also marked the split into two shells in the northwest and southeast.  
}
\label{fig:HB9compositeFigure}
\end{figure}

In Figure \ref{fig:HB9radioFigure}, I present radio images of HB9. The point-symmetric morphology is less prominent than in some other CCSNRs, and I identify it using the three radio images in Figure \ref{fig:HB9radioFigure} and Figure \ref{fig:HB9compositeFigure}.  
The most prominent structure protruding from the main SNR shell is the NW (northwest) ear. However, the radio image in Figure \ref{fig:HB9radioFigure} shows no opposite counterpart. The composite image of Figure \ref{fig:HB9compositeFigure} shows this ear to be a nozzle: an opening in the main SNR shell with edges extending outward. Moreover, this image shows a counter nozzle. I marked this pair of nozzles in Figure \ref{fig:HB9compositeFigure}. I attribute the shaping of this pair to a pair of two jets that participated in the explosion process in the JJEM framework. I mark the symmetry axis of this jet pair with a double-sided arrow, L1.   
%FFFFFFFFFFFFFFFFFFFFFFFFFFFFFFFFFFF
% 
%FFFFFFFFFFFFFFFFFFFFFFFFFFFFFFFFFFF
%FFFFFFFFFFFFFFFFFFFFFFFFFFFFFFFFFFF
% 
%FFFFFFFFFFFFFFFFFFFFFFFFFFFFFFFFFFF
    
A second prominent symmetry line is L2, which connects the N dent and the S dent; the S dent is on a bright bar. It is unclear whether this symmetry line is a jet axis, so I marked it with a dashed line. The dents suggest it is not the direction of a jet pair, while the bright arcs that extend along the same direction of this symmetry axis might indicate they were shaped by a pair of jets along this direction. In any case, this is a symmetry axis, part of the point-symmetric structure. A similar pair, best seen in panel (c), is along line L4, connecting the SW bar with the NE bar and its dent. 

The least prominent symmetry axis is L3. It connects the very small N ear (best seen in panel c) and the S ear, best seen in panel (c). I attribute the L3 symmetry axis to a pair of jets.  

A few words on the method I used to identify the point-symmetric morphology. 
This method follows the one widely used to characterize morphologies of planetary nebulae through careful visual inspection and then qualitative classification (e.g., \citealt{Balick1987, Parkeretal2006, Sahaietal2007, Kwok2024}; for a preliminary, more quantitative approach, see \citealt{ShishkinMichaelis2026}). This qualitative method has successfully led to the identification of jet-shaped structures  (e.g., \citealt{SahaiTrauger1998}), followed by comparison with hydrodynamical numerical simulations that revealed shaping processes in planetary nebulae (e.g., \citealt{GarciaSEguraetal2022, GarciaSEguraetal2025}), and in planetary nebulae and CCSNRs (e.g., \citealt{Akashietal2018}). 
This method of visual inspection revealed jet-shaped morphologies in other systems, such as B[e] supergiants (e.g., \citealt{Kashi2023}) and luminous blue variable nebulae (e.g., \citealt{Kashi2024}). The success of the visual inspection method, which might seem subjective, should not be a surprise even though it is not a quantitative one. This method has a large objective component. Humans have a remarkable ability to examine symmetry and critically identify departures from it by inspection alone. The reason is that facial symmetry is a key indicator of mate quality (e.g., \citealt{Rhodes2006, Pinheiroetal2023} and more references therein). Particularly, humans' ability to identify \textit{fluctuating asymmetries}, which are non-directional (random) deviations from perfect symmetry in bilaterally paired traits, is crucial, because fluctuating asymmetries tend to reflect health problems (e.g., \citealt{Rhodes2006}).  Recognition of symmetry is relatively automatic and consistent across cultures (e.g., \citealt{Rhodes2006}).  For the above arguments, I consider the identification of the point-symmetric morphology of SNR HB9 robust. 

I conclude that, although subtle, SNR HB9 has a point-symmetric morphology, composed of at least two pairs of jets that left their marks on CCSNR HB9. According to the JJEM, more jet pairs participated in the explosion; the others were probably weaker pairs of jets or occurred earlier in the explosion process.

% =========================
\section{The outer CO cloud and gamma ray emission  spots}
\label{sec:Extension}
% =========================

\begin{figure}
\centering
\includegraphics[trim=0.0cm 10.4cm 0cm 0.0cm, width=\textwidth]{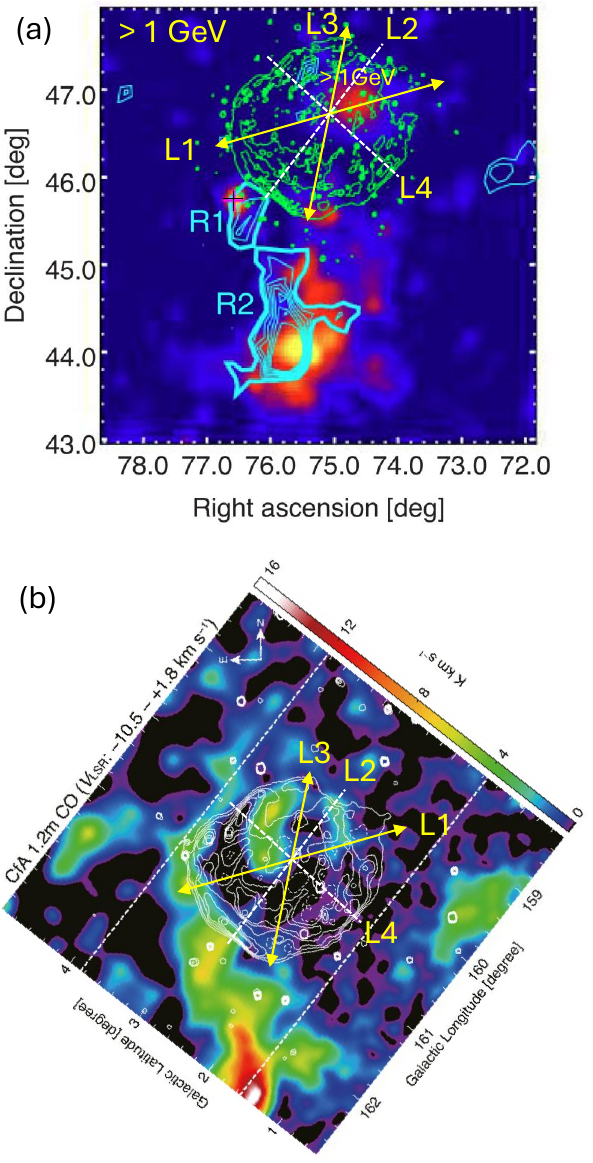}
\caption{
(a) A Fermi-LAT gamma-ray test statistic map in the vicinity of SNR HB9, adapted from \cite{Okaetal2022HB9}. The image shows emission at $>1~$GeV with the SNR main shell emission subtracted; the scale from low to high is: black, blue, red, yellow. 
The black-on-magenta cross indicates the position of PS J0506.5+4546.  
Green contours show the 1420 MHz radio emission, and the cyan contours south of the main SNR shell show the $^{12}$CO ($J = 1-0$) line intensity integrated over a velocity range between $-10.4$ and $+2.6 \km \s^{-1}$. I added the wind rose from Figures \ref{fig:HB9radioFigure} and  \ref{fig:HB9compositeFigure}. 
 (b) An image adapted from \cite{Sezeretal2019HB9} showing the intensity distributions of $^{12}$CO($J = 1-0$) in Galactic coordinates. The integration velocity is from $-10.5$ to $+1.8 \km \s^{-1}$. The white contours represent radio continuum intensity at 408 MHz. I rotated the image to match the orientation and scale of panel (a). I added the wind rose from Figures \ref{fig:HB9radioFigure} and  \ref{fig:HB9compositeFigure}. 
}
\label{Fig:HB9SouthFigure}
\end{figure}

To the south and southeast of the main shell of SNR HB9, there are spots of gamma-ray emission (e.g., \citealt{Okaetal2022HB9}). The upper panel of Figure \ref{Fig:HB9SouthFigure} (adapted from \citealt{Okaetal2022HB9}) presents these regions in red-yellow colors. These regions approximately coincide with regions of high-intensity $^{12}$CO ($J = 1-0$) emission (the cyan contours of panel a; the green contours on the main SNR shell represent radio emission). Panel (b) of Figure \ref{Fig:HB9SouthFigure} (adapted from \citealt{Sezeretal2019HB9}) presents the $^{12}$CO($J = 1 - 0$) emission in HB9 vicinity. \cite{Okaetal2022HB9} suggest that diffusion of protons from the SNR into the CO cloud leads to the gamma-ray emission south and southeast of the SNR HB9 main shell. 
% FFFFFFFFFFFFFFFFFFFFFFFFFFFFFFFFFFFFFFF
% FFFFFFFFFFFFFFFFFFFFFFFFFFFFFFFFFFFFFFF

I overlaid the wind rose from Figures \ref{fig:HB9radioFigure} and \ref{fig:HB9compositeFigure} on the two panels of Figure \ref{Fig:HB9SouthFigure}. I notice the following. (1) The CO cloud circles the SNR main shell from northeast to south. (2) The high-intensity $^{12}$CO($J = 1-0$) regions (yellow to red in panel b) close to the SNR main shell are along three of the axes, L1, L2, and L3. (3) The two bright gamma-ray emission spots (panel a) are along lines L2 and L3. 

Based on these properties, I suggest the following. (a) The protons from the SNR main shell that reach the CO cloud and give rise to the gamma-ray emission escape along openings in the main SNR shell that the jets opened along those directions. (b) I further speculate that fast material broke out of the main SNR shell, like the jets seen in some CCSNRs (e.g., Cassiopeia A and the Crab Nebula), and compressed CO molecular gas, forming dense spots. I leave this speculation for future hydrodynamical simulations.

% ==================================
\section{Jittering jets versus a CSM} 
\label{sec:CSM}
% ==================================

The explosion geometry, the circumstellar material (CSM) structure, and later the interstellar medium (ISM) structure influence SNR morphologies. 

In the case of SNR HB9, I attribute the point-symmetric morphology mainly to the explosion for the following reasons. (1) The CO cloud can influence only one side: northeast to south (panel b of Figure \ref{Fig:HB9SouthFigure}); it might have compressed the bright bar in the S dent + bar zone along line L2. (2) The two opposite nozzles along line L1 (Figure \ref{fig:HB9compositeFigure}) are compatible with opening from expanding jets from inside, and not from an interaction with an external medium that dents to compress gas. (3) In the northwest and southeast of SNR HB9, there are two arcs, which I term shells, as I mark in   Figure \ref{fig:HB9compositeFigure}. These shells are also visible in the radio maps, although less clearly. The two shells, inner and outer, have the same symmetry axis, L1.  The CSM and ISM cannot influence the inner shell. 

In other CCSNRs, there are other arguments for shaping by the explosion process rather than by the CSM or ISM (e.g., \citealt{SokerShishkin2025Vela}): The point-symmetric morphological features are enriched in metals from the inner core, like the jets in Cassiopeia A; a point symmetry in the inner regions unreachable to the CSM and ISM, as in the S-shaped structure of metals in the Vela SNR; the ejecta did not interact yet with the CSM, as the `Keyhole' structure of SN 1987A and its molecular gas (e.g., \citealt{Soker2026SN1987Amulecular}). 
A new indication for jet shaping comes from Doppler velocity maps of SNR S147 that I present here. 

SNR S147 (SNR G180.0-01.7) has two large prominent ears (e.g., \citealt{GrichenerSoker2017, Renetal2018RAA}).  While \cite{GrichenerSoker2017} and \cite{Shishkinetal2025S147} attributed the ears to pairs of jets during the explosion process, \cite{RenJJetal2026} argued recently that the two large ears of SNR S147 result from the supernova expansion into a CSM cavity. This cavity, they claim, had these ears from pre-explosion mass loss. 

\begin{figure}[th!]
\begin{center}
\includegraphics[trim=2.5cm 4.1cm 0cm 2.3cm, clip, width=0.55\textwidth]{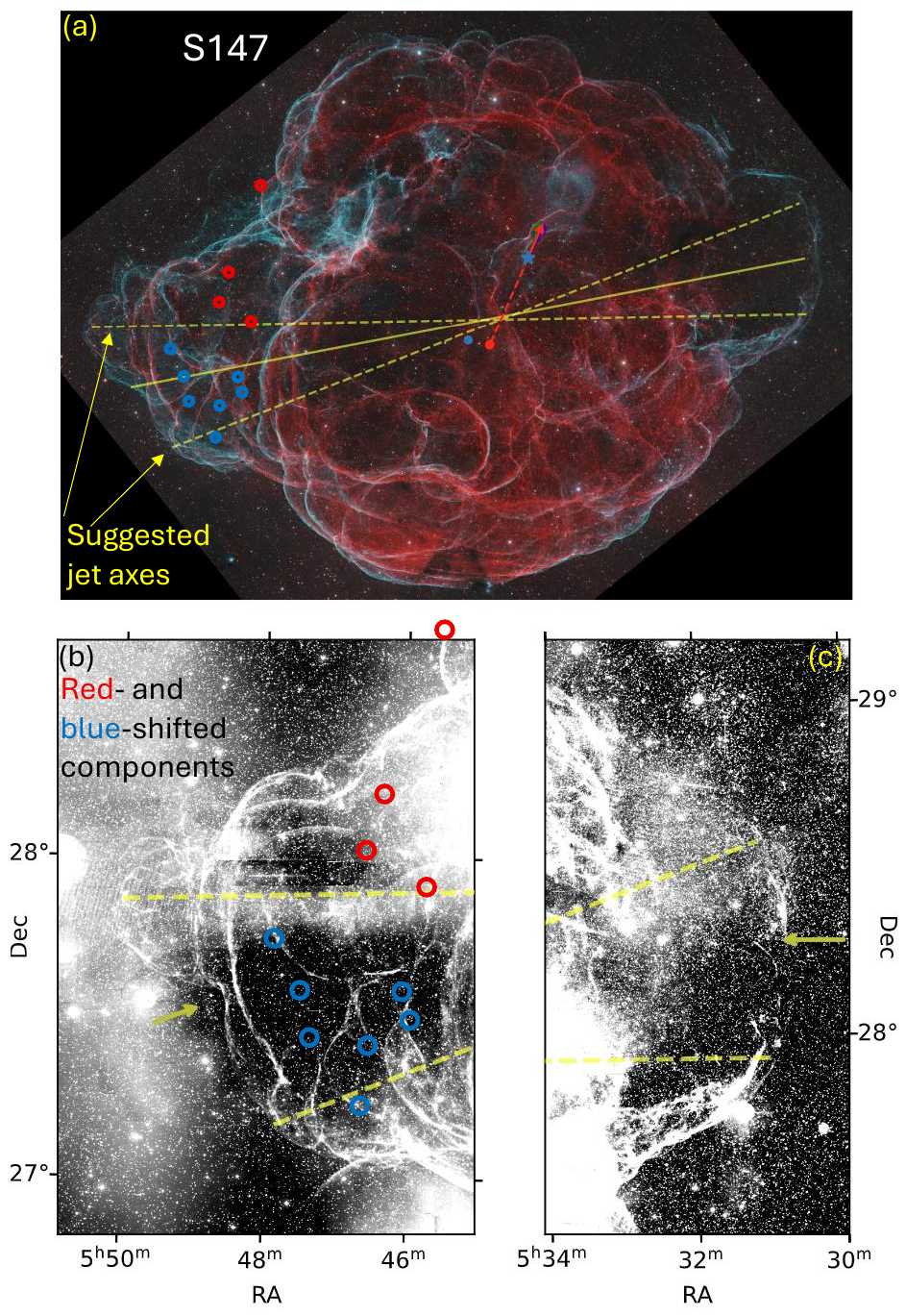} %[trim=left lower right upper]
\caption{A figure adapted from \cite{Shishkinetal2025S147} with their suggested two pairs of jets that shaped the pair of two wide ears during the explosion process. I added only in panels (a) and (b) the circles marking the fast redshifted (red) and blueshifted (blue) components of the eastern ear from \cite{RenJJetal2026}, as marked in Figure \ref{Fig:HB9DopplerFigure}.  (a) An Image of SNR S147 taken and processed by Mr. Christian Koll (\url{https://app.astrobin.com/i/vh6kx6}); blue and green represent [O \textsc{iii}], and red H$\alpha$. The two dashed yellow lines represent the proposed symmetry axes of two pairs of jets that shaped the two wide ears. A blue star is the location of the NS, while the blue dot is the geometric center of S147 as suggested by \citet{Kramer_etal_2003_center} based on radio observations. The yellow solid line between the two axes represents the main jet axis of S147. (b) Overexposed H$\alpha$ close-ups of the eastern ear. White smeared regions are artifacts of the overexposure. (c) As panel (b) for the western ear. }  
\label{fig:HB9DopplerJetFigure}
\end{center}
\end{figure}

\begin{figure}
\centering
\includegraphics[trim=0.0cm 5.0cm 0cm 0.0cm, width=0.70\textwidth]{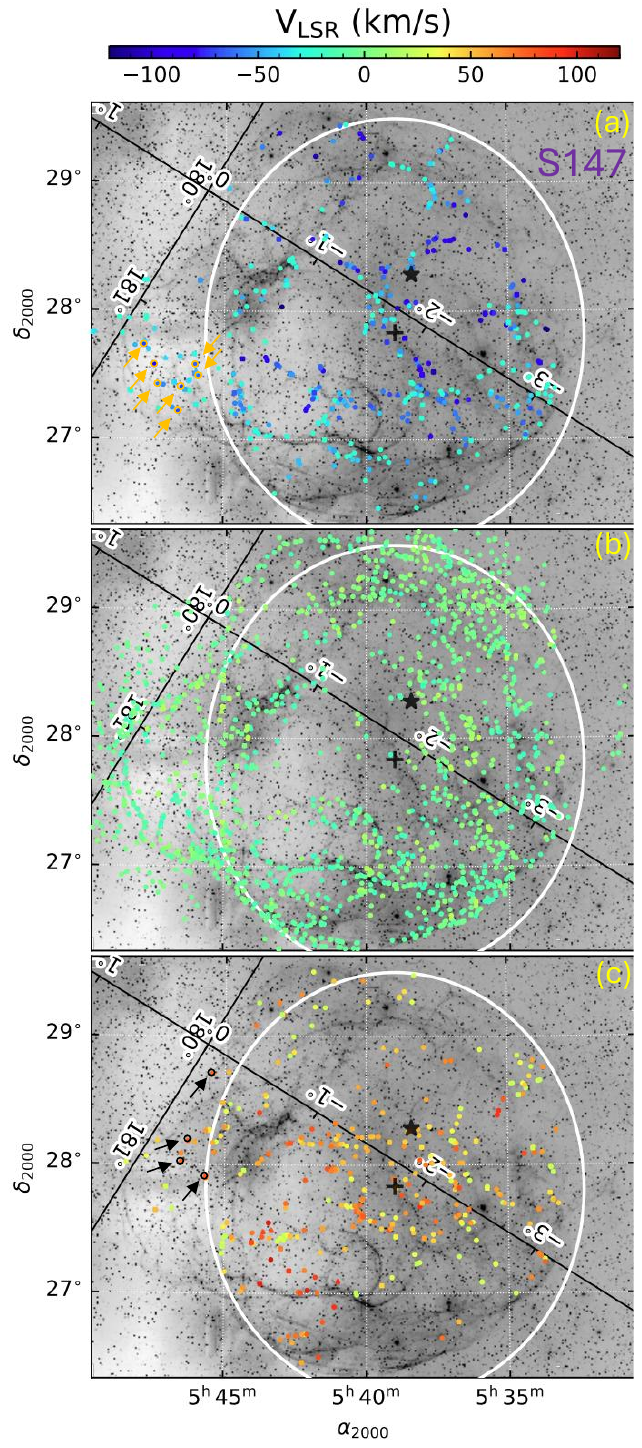}
\caption{ A figure adapted from \cite{RenJJetal2026} of the the spatial distribution of the Doppler shift velocity ${\rm V_{LSR}}$: (a) ${\rm V_{LSR}} < -20 \km \s^{-1}$, (b) $-20 \km \s^{-1} < {\rm V_{LSR}} < 20 \km \s^{-1}$, and (c) ${\rm V_{LSR}} > 20 \km \s^{-1}$. They overlaid an H$\alpha$ mosaic of S147 from the IGAPS survey. The black asterisk and cross mark the location of PSR J0538+2817 and the geometric center of S147 from \cite{Dinceletal2015}. 
They mark a 100 arcmin radius to define the main SNR shell.
I marked the 6 fastest blue-shifted components in the eastern ear (orange arrows and circles in panel a) and the 4 fastest red-shifted components (panel c, black arrows and circles); I copied these into Figure \ref{fig:HB9DopplerJetFigure}.   
}
\label{Fig:HB9DopplerFigure}
\end{figure}

Figure \ref{fig:HB9DopplerJetFigure}, which I adapted from \cite{Shishkinetal2025S147}, shows the morphology of S147, emphasizing the pair of opposite wide ears. \cite{Shishkinetal2025S147} proposed that two pairs of jets shaped this pair of ears; they marked the two pair axes with the dashed yellow lines. These jet pairs participated in the explosion processes of SNR S147 in the framework of the JJEM. The solid yellow line is drawn between these two axes to represent the main jet axis of S147.  
Figure \ref{Fig:HB9DopplerFigure}, adapted from \cite{RenJJetal2026}, shows the Doppler-shift velocities of many components in S147 as presented. I marked the fastest components in the eastern ears: six deep-blue colored components in panel (a), and four deep-red components in panel (c). I copied these locations to panels (a) and (b) of Figure \ref{fig:HB9DopplerJetFigure}.   
% FFFFFFFFFFFFFFFFFFFFFFFFFFFFFFFFFFFFFFFFFF
% FFFFFFFFFFFFFFFFFFFFFFFFFFFFFFFFFFFFFFFFFF
% FFFFFFFFFFFFFFFFFFFFFFFFFFFFFFFFFFFFFFF
% FFFFFFFFFFFFFFFFFFFFFFFFFFFFFFFFFFFFFFF

Panels (a) and (c) of Figure \ref{fig:HB9DopplerJetFigure} show that the red-shifted components that \cite{RenJJetal2026} identify in the eastern ear approximately coincide with the region that the northern jet that \cite{Shishkinetal2025S147} drew inflated, and the blue-shifted components in the eastern ear approximately coincide with the region inflated by the southern jet in that ear. I take it to support the claim that jets inflated the ears, rather than a cavity in the CSM. Namely, the eastern ear is composed of two separately inflated volumes: a red-shifted volume in the northern part and a blue-shifted volume in the southern part. This partition of the ear into two inflated volumes is compatible with the presence of the Prow only in the northern part of the ear. 

I conclude that the new velocity Doppler maps that \cite{RenJJetal2026} present support the claim that jets inflated the pair of ears of SNR 147. These two pairs of jets likely participated in the explosion process of S147 in the framework of the JJEM. 

That said, the CSM can also have a role in shaping SNRs. 
Outflows from massive stars are in many cases very complicated processes (e.g., \citealt{Michaelisetal2018, Gilkisetal2025, Grichener2025,  MukhijaKashi2025ApJ, MukhijaKashi2026ApJa, MukhijaKashi2026ApJb, MukhijaKashi2026NewA, MukhijaKashi2026PASP}), as the three CSM rings of SN 1987A demonstrate; these rings will definitely influence the morphology of SNR 1987A in the future. In some cases, jets shape the CSM (e.g., \citealt{AkashiKashi2020, Kashi2023}). So some morphological features might result from the CSM and ISM. However, I attribute the basic large-scale structure of the ears of S147 and of HB9 to jets that participated in the explosion mechanism. The CSM and ISM were likely to cause some departures from point-symmetry or have shaped less-prominent structural pairs.  

% ==================================
\section{Summary} 
\label{sec:Summary}
% ==================================

I identified a point-symmetric morphology in SNR HB9 (G160.9+2.6) using images from the literature (Figures \ref{fig:HB9radioFigure} and \ref{fig:HB9compositeFigure}). I identified four axes, L1-L4, as I marked in Figure \ref{fig:HB9radioFigure}. I attributed axes L1 and  L3, which connect pairs of ears, to two pairs of jets. Line L2 might also be an axis of a pair of jets. To identify the counterpart of the NW ear along line L1, I use a multi-wavelength image adapted from \cite{LiJetal2024HB9}, Figure \ref{fig:HB9compositeFigure}, which shows nozzles on both sides. Although the pairs of features of HB9 are not large, and hence the point-symmetric morphology is not as prominent as in some other CCSNRs, I consider the identification robust. The pairs of jets that shaped the identified point-symmetric morphology were part of the larger set of pairs of jets that exploded the progenitor of HB9 according to the JJEM.   

In Section \ref{sec:HB9}, I also pointed out that the visual inspection method I use, although qualitative, is quite objective and robust. It is heavily used to identify jet-shaped point-symmetric morphologies in cooling-flow clusters of galaxies (e.g., \citealt{Soker2024CF} for comparing CCSNRs and cooling flow clusters' X-ray morphologies) and planetary nebulae. 

In Section \ref{sec:Extension}, I examine the location of the bright gamma-ray emission regions to the south of the main SNR shell; these come from the interaction of protons escaping from the main SNR shell with the CO cloud (\citealt{Okaetal2022HB9}). Because the bright gamma-ray emission spots lie along some symmetry axes (Figure \ref{Fig:HB9SouthFigure}), I suggest that protons escaping from the main SNR shell do so along openings opened by the jets. I also speculate in Section \ref{sec:Extension} that jets that broke out of the main SNR shell compressed CO molecular gas to form denser CO emission spots. 

Undoubtedly, the CSM and ISM also influence the morphologies of old CCSNRs. In Section \ref{sec:CSM} I present arguments for the fact that the main point-symmetric structure of HB9 is due to pairs of jets, and not the CSM. Since the question of whether the shaping of point-symmetric structures observed in about 20 CCSNe is due to pairs of jets or the CSM, I further present one recent example, SNR S147. \cite{Shishkinetal2025S147} argued that two pairs of jets shaped the pair of wide ears of SNR S147. This is most prominent in the eastern ear, which is brighter and larger than the western ear (Figure \ref{fig:HB9DopplerJetFigure}). I used recently published Doppler velocity maps of the eastern ear of SNR S147 (Figure \ref{Fig:HB9DopplerFigure}) and found that the redshifted components approximately coincide with the volume the northern jet inflated according to \cite{Shishkinetal2025S147}, 
while the blueshifted components coincide with the volume inflated by the southern jet they proposed. I strengthen the earlier conclusions that at least two pairs of jets inflated the pair of wide ears of SNR S147. 

I marked some bright arcs on panel (b) of Figure \ref{fig:HB9radioFigure}. I cannot tell whether these are filaments on the surface of the SNR shell, the edges of three-dimensional shells or caps, or whether some are filaments and some are shells. Figure \ref{fig:HB9compositeFigure} suggests a double-shell structure in the northwest and southeast; the splitting into two shells (or a shell and a cap) occurs only in those directions, not across the entire sphere. Some other CCSNe possess such shells, e.g., SNR G309.2-00.6 and W44. The occurrence of two or three such shells has significant implications for the photosphere of CCSNe. During the explosion, when the outer shell becomes transparent, the photosphere moves to the slower inner shell/cap that still has an optical depth larger than one. Analysis of the photospheric velocity during the explosion will reveal two velocities, one after the other. 
\cite{SokerShiran2025} identified (in the observations by \citealt{Zimmermanetal2024}) two or possibly three photospheric shells in SN 2023ixf, and \cite{ShiranSoker2026} identified (based on observations by \citealt{ChenTWetal2025}) two photospheric shells in SN 2024ggi.   
SNR HB9 seems to add support to the common presence of two (or more) photospheric shells during the photospheric phase of CCSN explosions.   

My overall conclusion is that the point-symmetric morphology of SNR HB9, although subtle, is robust. As such, it supports the claim that the JJEM is the primary explosion mechanism for CCSNe and adds to the severe challenge for the neutrino-driven explosion mechanism to explain these morphologies. More accurately, I think the more than 20 point-symmetric CCSNRs rule out the neutrino-driven explosion mechanism (or limit it to be rare).    

% ===================================================
\section*{Acknowledgements}
% ===================================================

I thank the Charles Wolfson Academic Chair at the Technion for the support.
% BBBBBBBBBBBBBBBBBBBBBBBBBBBBBBBBBBBBBBBBBBB
\bibliography{BibReference}{}

@ARTICLE{PapishSoker2011,
       author = {{Papish}, Oded and {Soker}, Noam},
        title = "{Exploding core collapse supernovae with jittering jets}",
      journal = {\mnras},
         year = 2011,
        month = sep,
       volume = {416},
       number = {3},
        pages = {1697-1702},
          doi = {10.1111/j.1365-2966.2011.18671.x},
archivePrefix = {arXiv},
       eprint = {1103.1554},
 primaryClass = {astro-ph.HE},
       adsurl = {https://ui.adsabs.harvard.edu/abs/2011MNRAS.416.1697P}
}

@ARTICLE{Soker2010,
       author = {{Soker}, Noam},
        title = "{Applying the jet feedback mechanism to core-collapse supernova explosions}",
      journal = {\mnras},
         year = 2010,
        month = feb,
       volume = {401},
       number = {4},
        pages = {2793-2798},
          doi = {10.1111/j.1365-2966.2009.15862.x},
archivePrefix = {arXiv},
       eprint = {0909.5276},
 primaryClass = {astro-ph.HE},
       adsurl = {https://ui.adsabs.harvard.edu/abs/2010MNRAS.401.2793S}
}

@ARTICLE{Soker2022SNR0540,
       author = {{Soker}, Noam},
        title = "{Imprints of the Jittering Jets Explosion Mechanism in the Morphology of the Supernova Remnant SNR 0540-69.3}",
      journal = {Research in Astronomy and Astrophysics},
         year = 2022,
        month = mar,
       volume = {22},
       number = {3},
          eid = {035019},
        pages = {035019},
          doi = {10.1088/1674-4527/ac49e6},
archivePrefix = {arXiv},
       eprint = {2109.10230},
 primaryClass = {astro-ph.HE},
       adsurl = {https://ui.adsabs.harvard.edu/abs/2022RAA....22c5019S}
}

@ARTICLE{Soker2024NA1987A,
       author = {{Soker}, Noam},
        title = "{Hints of point-symmetric structures in SN 1987A: The jittering jets explosion mechanism}",
      journal = {\na},
         year = 2024,
        month = apr,
       volume = {107},
          eid = {102154},
        pages = {102154},
          doi = {10.1016/j.newast.2023.102154},
archivePrefix = {arXiv},
       eprint = {2309.07863},
 primaryClass = {astro-ph.HE},
       adsurl = {https://ui.adsabs.harvard.edu/abs/2024NewA..10702154S}
}

@ARTICLE{Soker2024PNSN,
       author = {{Soker}, Noam},
        title = "{Planetary Nebula Morphologies Indicate a Jet-Driven Explosion of SN 1987A and Other Core-Collapse Supernovae}",
      journal = {Galaxies},
         year = 2024,
        month = jun,
       volume = {12},
       number = {3},
          eid = {29},
        pages = {29},
          doi = {10.3390/galaxies12030029},
archivePrefix = {arXiv},
       eprint = {2404.14843},
 primaryClass = {astro-ph.HE},
       adsurl = {https://ui.adsabs.harvard.edu/abs/2024Galax..12...29S}
}

@ARTICLE{Soker2024CF,
       author = {{Soker}, Noam},
        title = "{Comparing jet-shaped point symmetry in cluster cooling flows and supernovae}",
      journal = {The Open Journal of Astrophysics},
         year = 2024,
        month = jun,
       volume = {7},
          eid = {49},
        pages = {49},
          doi = {10.33232/001c.120279},
archivePrefix = {arXiv},
       eprint = {2403.08544},
 primaryClass = {astro-ph.HE},
       adsurl = {https://ui.adsabs.harvard.edu/abs/2024OJAp....7...49S}
}

@ARTICLE{GrichenerSoker2017,
       author = {{Grichener}, Aldana and {Soker}, Noam},
        title = "{Core collapse supernova remnants with ears}",
      journal = {\mnras},
         year = 2017,
        month = jun,
       volume = {468},
       number = {1},
        pages = {1226-1235},
          doi = {10.1093/mnras/stx534},
archivePrefix = {arXiv},
       eprint = {1610.09647},
 primaryClass = {astro-ph.HE},
       adsurl = {https://ui.adsabs.harvard.edu/abs/2017MNRAS.468.1226G}
}

@ARTICLE{Soker2024Keyhole,
       author = {{Soker}, Noam},
        title = "{Supernova 1987A's Keyhole: A Long-lived Jet-pair in the Final Explosion Phase of Core-collapse Supernovae}",
      journal = {Research in Astronomy and Astrophysics},
         year = 2024,
        month = jul,
       volume = {24},
       number = {7},
          eid = {075006},
        pages = {075006},
          doi = {10.1088/1674-4527/ad4fc2},
archivePrefix = {arXiv},
       eprint = {2404.07455},
 primaryClass = {astro-ph.HE},
       adsurl = {https://ui.adsabs.harvard.edu/abs/2024RAA....24g5006S}
}

@ARTICLE{ShishkinSoker2025Crab,
       author = {{Shishkin}, Dmitry and {Soker}, Noam},
        title = "{Et tu, Brute?: The Crab Nebula also exploded by jittering jets}",
      journal = {arXiv e-prints},
         year = 2024,
        month = nov,
          eid = {arXiv:2411.07938},
        pages = {arXiv:2411.07938},
archivePrefix = {arXiv},
       eprint = {2411.07938},
 primaryClass = {astro-ph.HE},
       adsurl = {https://ui.adsabs.harvard.edu/abs/2024arXiv241107938S}
}

@ARTICLE{Soker2025Learning,
       author = {{Soker}, Noam},
        title = "{Learning from core-collapse supernova remnants on the explosion mechanism}",
      journal = {\na},
         year = 2025,
        month = dec,
       volume = {121},
          eid = {102453},
        pages = {102453},
          doi = {10.1016/j.newast.2025.102453},
archivePrefix = {arXiv},
       eprint = {2409.13657},
 primaryClass = {astro-ph.HE},
       adsurl = {https://ui.adsabs.harvard.edu/abs/2025NewA..12102453S}
}

@ARTICLE{Soker2024UnivReview,
       author = {{Soker}, Noam},
        title = "{The Two Alternative Explosion Mechanisms of Core-Collapse Supernovae: 2024 Status Report}",
      journal = {Universe},
         year = 2024,
        month = dec,
       volume = {10},
       number = {12},
          eid = {458},
        pages = {458},
          doi = {10.3390/universe10120458},
archivePrefix = {arXiv},
       eprint = {2411.08555},
 primaryClass = {astro-ph.HE},
       adsurl = {https://ui.adsabs.harvard.edu/abs/2024Univ...10..458S}
}

@ARTICLE{Shibataetal2025,
       author = {{Shibata}, Masaru and {Fujibayashi}, Sho and {Wanajo}, Shinya and {Ioka}, Kunihito and {Lam}, Alan Tsz-Lok and {Sekiguchi}, Yuichiro},
        title = "{Self-consistent scenario for jet and stellar explosions in collapsar: General relativistic magnetohydrodynamics simulation with a dynamo}",
      journal = {\prd},
         year = 2025,
        month = jun,
       volume = {111},
       number = {12},
          eid = {123017},
        pages = {123017},
          doi = {10.1103/msy2-fwhx},
archivePrefix = {arXiv},
       eprint = {2502.02077},
 primaryClass = {astro-ph.HE},
       adsurl = {https://ui.adsabs.harvard.edu/abs/2025PhRvD.111l3017S}
}

@ARTICLE{SokerShishkin2025Vela,
       author = {{Soker}, Noam and {Shishkin}, Dmitry},
        title = "{The Vela Supernova Remnant: The Unique Morphological Features of Jittering Jets}",
      journal = {Research in Astronomy and Astrophysics},
         year = 2025,
        month = mar,
       volume = {25},
       number = {3},
          eid = {035008},
        pages = {035008},
          doi = {10.1088/1674-4527/adb4cc},
archivePrefix = {arXiv},
       eprint = {2409.02626},
 primaryClass = {astro-ph.HE},
       adsurl = {https://ui.adsabs.harvard.edu/abs/2025RAA....25c5008S}
}

@ARTICLE{BearSoker2025,
       author = {{Bear}, Ealeal and {Soker}, Noam},
        title = "{Identifying a point-symmetric morphology in supernova remnant Cassiopeia A: Explosion by jittering jets}",
      journal = {\na},
         year = 2025,
        month = jan,
       volume = {114},
          eid = {102307},
        pages = {102307},
          doi = {10.1016/j.newast.2024.102307},
archivePrefix = {arXiv},
       eprint = {2403.07625},
 primaryClass = {astro-ph.HE},
       adsurl = {https://ui.adsabs.harvard.edu/abs/2025NewA..11402307B}
}

@ARTICLE{Sahaietal2007,
       author = {{Sahai}, Raghvendra and {Morris}, Mark and {S{\'a}nchez Contreras}, Carmen and {Claussen}, Mark},
        title = "{Preplanetary Nebulae: A Hubble Space Telescope Imaging Survey and a New Morphological Classification System}",
      journal = {\aj},
         year = 2007,
        month = dec,
       volume = {134},
       number = {6},
        pages = {2200-2225},
          doi = {10.1086/522944},
archivePrefix = {arXiv},
       eprint = {0707.4662},
 primaryClass = {astro-ph},
       adsurl = {https://ui.adsabs.harvard.edu/abs/2007AJ....134.2200S}
}

@ARTICLE{Kramer_etal_2003_center,
       author = {{Kramer}, M. and {Lyne}, A.~G. and {Hobbs}, G. and {L{\"o}hmer}, O. and {Carr}, P. and {Jordan}, C. and {Wolszczan}, A.},
        title = "{The Proper Motion, Age, and Initial Spin Period of PSR J0538+2817 in S147}",
      journal = {\apjl},
         year = 2003,
        month = aug,
       volume = {593},
       number = {1},
        pages = {L31-L34},
          doi = {10.1086/378082},
archivePrefix = {arXiv},
       eprint = {astro-ph/0306628},
 primaryClass = {astro-ph},
       adsurl = {https://ui.adsabs.harvard.edu/abs/2003ApJ...593L..31K}
}

@ARTICLE{Braudoetal2025,
       author = {{Braudo}, Jessica and {Michaelis}, Amir and {Akashi}, Muhammad and {Soker}, Noam},
        title = "{Simulating the Shaping of Point-symmetric Structures in the Jittering Jets Explosion Mechanism}",
      journal = {\pasp},
         year = 2025,
        month = may,
       volume = {137},
       number = {5},
          eid = {054201},
        pages = {054201},
          doi = {10.1088/1538-3873/add08e},
archivePrefix = {arXiv},
       eprint = {2503.10326},
 primaryClass = {astro-ph.HE},
       adsurl = {https://ui.adsabs.harvard.edu/abs/2025PASP..137e4201B}
}

@ARTICLE{Dinceletal2015,
       author = {{Din{\c{c}}el}, B. and {Neuh{\"a}user}, R. and {Yerli}, S.~K. and {Ankay}, A. and {Tetzlaff}, N. and {Torres}, G. and {Mugrauer}, M.},
        title = "{Discovery of an OB runaway star inside SNR S147}",
      journal = {\mnras},
         year = 2015,
        month = apr,
       volume = {448},
       number = {4},
        pages = {3196-3205},
          doi = {10.1093/mnras/stv124},
archivePrefix = {arXiv},
       eprint = {1501.07220},
 primaryClass = {astro-ph.SR},
       adsurl = {https://ui.adsabs.harvard.edu/abs/2015MNRAS.448.3196D}
}

@ARTICLE{Rhodes2006,
       author = {{Rhodes}, Gillianr},
        title = "{THE EVOLUTIONARY PSYCHOLOGY OF FACIAL BEAUTY}",
      journal = {Annu. Rev. Psychol},
         year = 2006,
        month = Jan,
       volume = {57},
       number = {2},
          eid = {199},
        pages = {199},
          doi = {10.1146/annurev.psych.57.102904.190208},
archivePrefix = {arXiv},
       eprint = {0000.000000},
 primaryClass = {astro-ph.HE},
       adsurl = {}
}

@ARTICLE{Pinheiroetal2023,
       author = {{Pinheiro}, Luiza Penha and {Monteiro}, Luis Carlos Pereira and {Henriques}, Leonardo Dutra and {Souza}, Givago Silva and {Miranda}, Ana Catarina and {Costa}, Marcelo Fernandes and {Henriques}, Alda Loureiro},
        title = "{Association between Facial Metrics and Mate Rejection for Long-Term Relationship by Heterosexual Men}",
      journal = {Symmetry},
         year = 2023,
        month = jan,
       volume = {15},
       number = {1},
          eid = {133},
        pages = {133},
          doi = {10.3390/sym15010133},
       adsurl = {https://ui.adsabs.harvard.edu/abs/2023Symm...15..133P}
}

@ARTICLE{BetheWilson1985_OGdelayedneutrino,
       author = {{Bethe}, H.~A. and {Wilson}, J.~R.},
        title = "{Revival of a stalled supernova shock by neutrino heating}",
      journal = {\apj},
         year = 1985,
        month = aug,
       volume = {295},
        pages = {14-23},
          doi = {10.1086/163343},
       adsurl = {https://ui.adsabs.harvard.edu/abs/1985ApJ...295...14B}
}

@ARTICLE{Shishkinetal2025S147,
       author = {{Shishkin}, Dmitry and {Bear}, Ealeal and {Soker}, Noam},
        title = "{Natal Kick by Early-asymmetrical Pairs of Jets to the Neutron Star of Supernova Remnant S147}",
      journal = {\apj},
         year = 2025,
        month = oct,
       volume = {992},
       number = {2},
          eid = {190},
        pages = {190},
          doi = {10.3847/1538-4357/ae0332},
archivePrefix = {arXiv},
       eprint = {2506.21548},
 primaryClass = {astro-ph.HE},
       adsurl = {https://ui.adsabs.harvard.edu/abs/2025ApJ...992..190S}
}

@ARTICLE{Gilkisetal2025,
       author = {{Gilkis}, Avishai and {Laplace}, Eva and {Arcavi}, Iair and {Shenar}, Tomer and {Schneider}, Fabian R.~N.},
        title = "{The landscape of binary core-collapse supernova progenitors and the late emergence of Wolf{\textendash}Rayet winds}",
      journal = {\mnras},
         year = 2025,
        month = jul,
       volume = {540},
       number = {4},
        pages = {3094-3120},
          doi = {10.1093/mnras/staf884},
archivePrefix = {arXiv},
       eprint = {2503.01993},
 primaryClass = {astro-ph.SR},
       adsurl = {https://ui.adsabs.harvard.edu/abs/2025MNRAS.540.3094G}
}

@ARTICLE{WangShishkinSoker2026,
       author = {{Wang}, Nikki Yat Ning and {Shishkin}, Dmitry and {Soker}, Noam},
        title = "{Jittering jets in stripped-envelope core-collapse supernovae}",
      journal = {arXiv e-prints},
         year = 2025,
        month = oct,
          eid = {arXiv:2510.02203},
        pages = {arXiv:2510.02203},
          doi = {10.48550/arXiv.2510.02203},
archivePrefix = {arXiv},
       eprint = {2510.02203},
 primaryClass = {astro-ph.HE},
       adsurl = {https://ui.adsabs.harvard.edu/abs/2025arXiv251002203W}
}

@ARTICLE{SahaiTrauger1998,
       author = {{Sahai}, Raghvendra and {Trauger}, John T.},
        title = "{Multipolar Bubbles and Jets in Low-Excitation Planetary Nebulae: Toward a New Understanding of the Formation and Shaping of Planetary Nebulae}",
      journal = {\aj},
         year = 1998,
        month = sep,
       volume = {116},
       number = {3},
        pages = {1357-1366},
          doi = {10.1086/300504},
       adsurl = {https://ui.adsabs.harvard.edu/abs/1998AJ....116.1357S}
}

@ARTICLE{Parkeretal2006,
       author = {{Parker}, Quentin A. and {Acker}, A. and {Frew}, D.~J. and {Hartley}, M. and {Peyaud}, A.~E.~J. and {Ochsenbein}, F. and {Phillipps}, S. and {Russeil}, D. and {Beaulieu}, S.~F. and {Cohen}, M. and {K{\"o}ppen}, J. and {Miszalski}, B. and {Morgan}, D.~H. and {Morris}, R.~A.~H. and {Pierce}, M.~J. and {Vaughan}, A.~E.},
        title = "{The Macquarie/AAO/Strasbourg H{\ensuremath{\alpha}} Planetary Nebula Catalogue: MASH}",
      journal = {\mnras},
         year = 2006,
        month = nov,
       volume = {373},
       number = {1},
        pages = {79-94},
          doi = {10.1111/j.1365-2966.2006.10950.x},
       adsurl = {https://ui.adsabs.harvard.edu/abs/2006MNRAS.373...79P}
}

@ARTICLE{Balick1987,
       author = {{Balick}, Bruce},
        title = "{The Evolution of Planetary Nebulae. I. Structures, Ionizations, and Morphological Sequences}",
      journal = {\aj},
         year = 1987,
        month = sep,
       volume = {94},
        pages = {671},
          doi = {10.1086/114504},
       adsurl = {https://ui.adsabs.harvard.edu/abs/1987AJ.....94..671B}
}

@ARTICLE{GarciaSEguraetal2022,
       author = {{Garc{\'\i}a-Segura}, Guillermo and {Taam}, Ronald E. and {Ricker}, Paul M.},
        title = "{Common-envelope shaping of planetary nebulae - IV. From protoplanetary to planetary nebula}",
      journal = {\mnras},
         year = 2022,
        month = dec,
       volume = {517},
       number = {3},
        pages = {3822-3831},
          doi = {10.1093/mnras/stac2824},
archivePrefix = {arXiv},
       eprint = {2209.15081},
 primaryClass = {astro-ph.SR},
       adsurl = {https://ui.adsabs.harvard.edu/abs/2022MNRAS.517.3822G}
}

@ARTICLE{GarciaSEguraetal2025,
       author = {{Garc{\'\i}a-Segura}, Guillermo and {Manchado}, Arturo and {Toal{\'a}}, Jes{\'u}s A. and {Guerrero}, Mart{\'\i}n A. and {Castro-Tirado}, Alberto J.},
        title = "{Planetary nebula evolution for single stellar models. The formation of neutral spikes}",
      journal = {\mnras},
         year = 2025,
        month = nov,
       volume = {543},
       number = {4},
        pages = {3867-3884},
          doi = {10.1093/mnras/staf1744},
archivePrefix = {arXiv},
       eprint = {2510.08721},
 primaryClass = {astro-ph.SR},
       adsurl = {https://ui.adsabs.harvard.edu/abs/2025MNRAS.543.3867G}
}

@ARTICLE{ShiranSoker2026,
       author = {{Shiran}, Kobi and {Soker}, Noam},
        title = "{Double shell structure in supernova 2024ggi}",
      journal = {arXiv e-prints},
         year = 2026,
        month = dec,
          eid = {arXiv:2512.16554},
        pages = {arXiv:2512.16554},
archivePrefix = {arXiv},
       eprint = {2512.16554},
 primaryClass = {astro-ph.HE},
       adsurl = {https://ui.adsabs.harvard.edu/abs/2025arXiv251216554S}
}

@ARTICLE{Soker2025G11,
       author = {{Soker}, Noam},
        title = "{Point-symmetric morphology in supernova remnant G11.2-0.3: the jittering jets explosion mechanism}",
      journal = {arXiv e-prints},
         year = 2025,
        month = nov,
          eid = {arXiv:2511.14578},
        pages = {arXiv:2511.14578},
          doi = {10.48550/arXiv.2511.14578},
archivePrefix = {arXiv},
       eprint = {2511.14578},
 primaryClass = {astro-ph.HE},
       adsurl = {https://ui.adsabs.harvard.edu/abs/2025arXiv251114578S}
}

@ARTICLE{ShishkinMichaelis2026,
       author = {{Shishkin}, Dmitry and {Michaelis}, Amir},
        title = "{Quantifying Symmetry: Transformation Information for Planetary Nebulae and Supernova Remnants}",
      journal = {\pasp},
         year = 2026,
        month = jun,
       volume = {138},
       number = {6},
          eid = {064508},
        pages = {064508},
          doi = {10.1088/1538-3873/ae7692},
       adsurl = {https://ui.adsabs.harvard.edu/abs/2026PASP..138f4508S}
}

@ARTICLE{Renetal2018RAA,
       author = {{Ren}, Juan-Juan and {Liu}, Xiao-Wei and {Chen}, Bing-Qiu and {Xiang}, Mao-Sheng and {Yuan}, Hai-Bo and {Huang}, Yang and {Zhang}, Hua-Wei and {Wang}, Chun and {Tian}, Zhi-Jia and {Liu}, Gao-Chao and {Wu}, Hong},
        title = "{Mapping the emission line strengths and kinematics of supernova remnant S147 with extensive LAMOST spectroscopic observations}",
      journal = {Research in Astronomy and Astrophysics},
         year = 2018,
        month = aug,
       volume = {18},
       number = {9},
          eid = {111},
        pages = {111},
          doi = {10.1088/1674-4527/18/9/111},
archivePrefix = {arXiv},
       eprint = {1804.10989},
 primaryClass = {astro-ph.GA},
       adsurl = {https://ui.adsabs.harvard.edu/abs/2018RAA....18..111R}
}

@ARTICLE{PanLi2026,
       author = {{Pan}, Kuo-Chuan and {Li}, Yi-Fang},
        title = "{From Jets to Failed Supernovae: Morphologies and Gravitational-Wave Signatures in Two-Dimensional Magnetorotational Core-Collapse Supernovae}",
      journal = {arXiv e-prints},
         year = 2026,
        month = mar,
          eid = {arXiv:2603.25846},
        pages = {arXiv:2603.25846},
archivePrefix = {arXiv},
       eprint = {2603.25846},
 primaryClass = {astro-ph.HE},
       adsurl = {https://ui.adsabs.harvard.edu/abs/2026arXiv260325846P}
}

@ARTICLE{Akashietal2018,
       author = {{Akashi}, Muhammad and {Bear}, Ealeal and {Soker}, Noam},
        title = "{Forming H-shaped and barrel-shaped nebulae with interacting jets}",
      journal = {\mnras},
         year = 2018,
        month = apr,
       volume = {475},
       number = {4},
        pages = {4794-4808},
          doi = {10.1093/mnras/sty029},
archivePrefix = {arXiv},
       eprint = {1712.07156},
 primaryClass = {astro-ph.SR},
       adsurl = {https://ui.adsabs.harvard.edu/abs/2018MNRAS.475.4794A}
}

@ARTICLE{SokerShiran2025,
       author = {{Soker}, Noam and {Shiran}, Kobi},
        title = "{Multiple Shells in Supernova 2023ixf Support the Jittering Jets Explosion Mechanism}",
      journal = {\apjl},
         year = 2025,
        month = dec,
       volume = {995},
       number = {1},
          eid = {L3},
        pages = {L3},
          doi = {10.3847/2041-8213/ae21da},
archivePrefix = {arXiv},
       eprint = {2510.18782},
 primaryClass = {astro-ph.HE},
       adsurl = {https://ui.adsabs.harvard.edu/abs/2025ApJ...995L...3S}
}

@ARTICLE{Zimmermanetal2024,
       author = {{Zimmerman}, E.~A. and {Irani}, I. and {Chen}, P. and {Gal-Yam}, A. and {Schulze}, S. and {Perley}, D.~A. and {Sollerman}, J. and {Filippenko}, A.~V. and {Shenar}, T. and {Yaron}, O. and {Shahaf}, S. and {Bruch}, R.~J. and {Ofek}, E.~O. and {De Cia}, A. and {Brink}, T.~G. and {Yang}, Y. and {Vasylyev}, S.~S. and {Ben Ami}, S. and {Aubert}, M. and {Badash}, A. and {Bloom}, J.~S. and {Brown}, P.~J. and {De}, K. and {Dimitriadis}, G. and {Fransson}, C. and {Fremling}, C. and {Hinds}, K. and {Horesh}, A. and {Johansson}, J.~P. and {Kasliwal}, M.~M. and {Kulkarni}, S.~R. and {Kushnir}, D. and {Martin}, C. and {Matuzewski}, M. and {McGurk}, R.~C. and {Miller}, A.~A. and {Morag}, J. and {Neil}, J.~D. and {Nugent}, P.~E. and {Post}, R.~S. and {Prusinski}, N.~Z. and {Qin}, Y. and {Raichoor}, A. and {Riddle}, R. and {Rowe}, M. and {Rusholme}, B. and {Sfaradi}, I. and {Sjoberg}, K.~M. and {Soumagnac}, M. and {Stein}, R.~D. and {Strotjohann}, N.~L. and {Terwel}, J.~H. and {Wasserman}, T. and {Wise}, J. and {Wold}, A. and {Yan}, L. and {Zhang}, K.},
        title = "{The complex circumstellar environment of supernova 2023ixf}",
      journal = {\nat},
         year = 2024,
        month = mar,
       volume = {627},
       number = {8005},
        pages = {759-762},
          doi = {10.1038/s41586-024-07116-6},
archivePrefix = {arXiv},
       eprint = {2310.10727},
 primaryClass = {astro-ph.HE},
       adsurl = {https://ui.adsabs.harvard.edu/abs/2024Natur.627..759Z}
}

@ARTICLE{ChenTWetal2025,
       author = {{Chen}, Ting-Wan and {Yang}, Sheng and {Srivastav}, Shubham and {Moriya}, Takashi J. and {Smartt}, Stephen J. and {Rest}, Sofia and {Rest}, Armin and {Lin}, Hsing Wen and {Miao}, Hao-Yu and {Cheng}, Yu-Chi and {Aryan}, Amar and {Cheng}, Chia-Yu and {Fraser}, Morgan and {Huang}, Li-Ching and {Lee}, Meng-Han and {Lai}, Cheng-Han and {Liu}, Yu-Hsuan and {Sankar. K}, Aiswarya and {Smith}, Ken W. and {Stevance}, Heloise F. and {Wang}, Ze-Ning and {Anderson}, Joseph P. and {Angus}, Charlotte R. and {de Boer}, Thomas and {Chambers}, Kenneth and {Duan}, Hao-Yuan and {Erasmus}, Nicolas and {Fulton}, Michael and {Gao}, Hua and {Herman}, Joanna and {Hou}, Wei-Jie and {Hsiao}, Hsiang-Yao and {Huber}, Mark E. and {Lin}, Chien-Cheng and {Lin}, Hung-Chin and {Magnier}, Eugene A. and {Man}, Ka Kit and {Moore}, Thomas and {Ngeow}, Chow-Choong and {Nicholl}, Matt and {Ou}, Po-Sheng and {Pignata}, Giuliano and {Shiau}, Yu-Chien and {Sommer}, Julian Silvester and {Tonry}, John L. and {Wang}, Xiao-Feng and {Wainscoat}, Richard and {Young}, David R. and {Yeh}, You-Ting and {Zhang}, Jujia},
        title = "{Discovery and Extensive Follow-up of SN 2024ggi, a Nearby Type IIP Supernova in NGC 3621}",
      journal = {\apj},
         year = 2025,
        month = apr,
       volume = {983},
       number = {1},
          eid = {86},
        pages = {86},
          doi = {10.3847/1538-4357/adb428},
archivePrefix = {arXiv},
       eprint = {2406.09270},
 primaryClass = {astro-ph.HE},
       adsurl = {https://ui.adsabs.harvard.edu/abs/2025ApJ...983...86C}
}

@ARTICLE{Giudicietal2026,
       author = {{Giudici}, Beatrice and {Gabler}, Michael and {Janka}, Hans-Thomas},
        title = "{Hydrodynamic instabilities in long-term three-dimensional simulations of neutrino-driven supernovae of 13 red supergiant progenitors}",
      journal = {arXiv e-prints},
         year = 2025,
        month = nov,
          eid = {arXiv:2511.11796},
        pages = {arXiv:2511.11796},
archivePrefix = {arXiv},
       eprint = {2511.11796},
 primaryClass = {astro-ph.HE},
       adsurl = {https://ui.adsabs.harvard.edu/abs/2025arXiv251111796G}
}

@ARTICLE{LuoZhaKajino2026,
       author = {{Luo}, Y. and {Zha}, S. and {Kajino}, T.},
        title = "{Systematic study of the impact of magnetic fields on neutrino transport in core-collapse supernovae}",
      journal = {\prd},
         year = 2026,
        month = jan,
       volume = {113},
       number = {2},
          eid = {023024},
        pages = {023024},
          doi = {10.1103/7ytg-wzl8},
archivePrefix = {arXiv},
       eprint = {2512.10417},
 primaryClass = {astro-ph.HE},
       adsurl = {https://ui.adsabs.harvard.edu/abs/2026PhRvD.113b3024L}
}

@ARTICLE{Wessonetal2026,
       author = {{Wesson}, R. and {Gabler}, M. and {Lyons}, M. and {Wildman}, J. and {Matsuura}, Mikako and {Janka}, H.-T. and {Giudici}, B. and {Cigan}, P. and {Gomez}, H.~L. and {Indebetouw}, R. and {Richards}, A.~M.~S. and {Wongwathanarat}, A.},
        title = "{3D insights into SN 1987A: ALMA observations compared to hydrodynamical explosion simulations}",
      journal = {\mnras},
         year = 2026,
        month = may,
       volume = {548},
       number = {2},
          eid = {stag325},
        pages = {stag325},
          doi = {10.1093/mnras/stag325},
archivePrefix = {arXiv},
       eprint = {2602.11259},
 primaryClass = {astro-ph.HE},
       adsurl = {https://ui.adsabs.harvard.edu/abs/2026MNRAS.548ag325W}
}

@ARTICLE{Mannoetal2026,
       author = {{Manno}, Marco and {Mart{\'\i}nez-Mirav{\'e}}, Pablo and {Tamborra}, Irene},
        title = "{Matter- and magnetically-driven flavor conversion of neutrinos in magnetorotational collapses}",
      journal = {Classical and Quantum Gravity},
         year = 2026,
        month = jun,
       volume = {43},
       number = {11},
          eid = {115001},
        pages = {115001},
          doi = {10.1088/1361-6382/ae6f69},
archivePrefix = {arXiv},
       eprint = {2603.08843},
 primaryClass = {astro-ph.HE},
       adsurl = {https://ui.adsabs.harvard.edu/abs/2026CQGra..43k5001M}
}

@ARTICLE{BoppGottlieb2025,
       author = {{Bopp}, Justin and {Gottlieb}, Ore},
        title = "{Fast Transients from Magnetic Disks around Nonspinning Collapsar Black Holes}",
      journal = {\apjl},
         year = 2025,
        month = apr,
       volume = {982},
       number = {2},
          eid = {L56},
        pages = {L56},
          doi = {10.3847/2041-8213/adbdcd},
archivePrefix = {arXiv},
       eprint = {2410.22401},
 primaryClass = {astro-ph.HE},
       adsurl = {https://ui.adsabs.harvard.edu/abs/2025ApJ...982L..56B}
}

@ARTICLE{Soker2026SN1987Amulecular,
       author = {{Soker}, Noam},
        title = "{Molecular bipolar outflow in SN 1987A supports the jittering-jets explosion mechanism}",
      journal = {arXiv e-prints},
         year = 2026,
        month = feb,
          eid = {arXiv:2602.21004},
        pages = {arXiv:2602.21004},
          doi = {10.48550/arXiv.2602.21004},
archivePrefix = {arXiv},
       eprint = {2602.21004},
 primaryClass = {astro-ph.HE},
       adsurl = {https://ui.adsabs.harvard.edu/abs/2026arXiv260221004S}
}

@ARTICLE{Soker2026Failed,
       author = {{Soker}, Noam},
        title = "{The failed failed-supernova scenario of M31-2014-DS1}",
      journal = {arXiv e-prints},
         year = 2026,
        month = jan,
          eid = {arXiv:2601.14497},
        pages = {arXiv:2601.14497},
          doi = {10.48550/arXiv.2601.14497},
archivePrefix = {arXiv},
       eprint = {2601.14497},
 primaryClass = {astro-ph.HE},
       adsurl = {https://ui.adsabs.harvard.edu/abs/2026arXiv260114497S}
}

@ARTICLE{AkashiSoker2026a,
       author = {{Akashi}, Muhammad and {Soker}, Noam},
        title = "{Simulating the jittering-jets explosion mechanism: circum-jet rings account for observed core-collapse supernova remnant morphologies}",
      journal = {arXiv e-prints},
         year = 2026,
        month = mar,
          eid = {arXiv:2603.29527},
        pages = {arXiv:2603.29527},
archivePrefix = {arXiv},
       eprint = {2603.29527},
 primaryClass = {astro-ph.HE},
       adsurl = {https://ui.adsabs.harvard.edu/abs/2026arXiv260329527A}
}

@ARTICLE{AkashiSoker2026BG11,
       author = {{Akashi}, Muhammad and {Soker}, Noam},
        title = "{Reproducing Morphological Features in the Supernova Remnant G11.2-0.3 by Simulating Jittering Jets}",
      journal = {\pasp},
         year = 2026,
        month = jun,
       volume = {138},
       number = {6},
          eid = {064203},
        pages = {064203},
          doi = {10.1088/1538-3873/ae7903},
archivePrefix = {arXiv},
       eprint = {2605.12356},
 primaryClass = {astro-ph.HE},
       adsurl = {https://ui.adsabs.harvard.edu/abs/2026PASP..138f4203A}
}

@ARTICLE{Soker2026DustJets,
       author = {{Soker}, Noam},
        title = "{Jittering Jets Promote Dust Formation in Core-Collapse Supernovae}",
      journal = {International Journal of Astronomy and Astrophysics},
         year = 2026,
        month = jan,
       volume = {16},
       number = {2},
        pages = {112-125},
          doi = {10.4236/ijaa.2026.162008},
archivePrefix = {arXiv},
       eprint = {2509.19264},
 primaryClass = {astro-ph.HE},
       adsurl = {https://ui.adsabs.harvard.edu/abs/2026IJAA...16..112S}
}

@ARTICLE{Akaho2026,
       author = {{Akaho}, Ryuichiro},
        title = "{Momentum Space Resolution Dependence in Boltzmann Neutrino Radiation Hydrodynamics Simulations of Core-collapse Supernovae}",
      journal = {\mnras},
         year = 2026,
        month = sep,
          doi = {10.1093/mnras/stag1677},
archivePrefix = {arXiv},
       eprint = {2607.12891},
 primaryClass = {astro-ph.HE},
       adsurl = {https://ui.adsabs.harvard.edu/abs/2026MNRAS.tmp.1556A}
}
\bibliographystyle{aasjournal}
% BBBBBBBBBBBBBBBBBBBBBBBBBBBBBBBBBBBBBBBBBBB

\end{document}